# A null model for testing thermodynamic optimization in ecological systems

Santiago R. Doyle[1,2,3], Florencia Carusela[1,2], Sebastián Guala[1] &

Fernando Momo[1,3]

1: Instituto de Ciencias, Universidad Nacional de General Sarmiento, J.M. Gutierrez 1150, 1613 Los Polvorines, Argentina

2: Consejo Nacional de Investigaciones Científicas y Técnicas, Rivadavia 1917, 1033 Ciudad Autónoma de Buenos Aires, Argentina

3: Programa de Investigaciones en Ecología Acuática, Departamento de Ciencias Básicas, Universidad Nacional de Luján, C.C. 221, 6700 Luján, Argentina

E-mail: *sdoyle@ungs.edu.ar*

Abstract

Several authors have hypothesized that ecological systems are subject to thermodynamic optimization, which, if proven correct, could represent a long sought general principle of organization in ecology. Although there have been recent advances, this still remains as an unresolved topic, and ecologists lack a general method to test thermodynamic optimization hypotheses in specific systems. Here we present a general, novel approach that allows generating a null model for testing thermodynamic optimization on ecological systems. We first describe the general methodology, which is based in the analysis of a parametrized mathematical model of the system and the explicit consideration of constraints. Next we present an application example to an animal population using a general age-structured population model and physiological parameters from the literature. We finalize discussing the relevance of this work in the context of the current state of ecology, and implications for the further development of a thermodynamic ecological theory.

**Introduction**

Since the beginning of ecology, general principles that explain the steady state of ecological systems have been sought. Although nowadays there are many examples of successful theories that explain various ecological patterns, the notion that ecology as a whole lacks a general framework that can turn different theories into a coherent scheme have led to the debate of whether there are general laws in ecology (Lawton 1999; Simberloff 2004; Lange 2005; Scheiner & Willig 2008). Several authors have argued that a thermodynamic based approach to ecology represents a promising framework to unify apparently dissimilar theories into general principles, describing patterns and processes in terms of relevant thermodynamic properties (e.g. Ernest et al. 2003; Brown et al. 2004; Jørgensen & Svirezhev 2004; Meysman & Bruers 2010). However, there is currently no agreement regarding to what extent properties of actual ecological systems can be predicted based on thermodynamic considerations (Dewar 2010) mas and the search for a thermodynamic-based ecological theory is still an open quest.

Lotka (1922a) was the first to propose a thermodynamic organizing principle to which ecological systems would be subject to, describing what was afterward referred as the maximum power principle (Fath et al. 2001). Lotka reasoned that if energy supply is limited, individuals that attain higher rates of energy acquisition would have an advantage over competing individuals, and thus the energy acquisition rate (power) would tend to a maximum by means of natural selection. Lotka hypothesized that ecological systems of all hierarchies would be subject to this thermodynamic principle, on the grounds that he considered it as a physical rather than biological law (Lotka 1922b). Since the seminal work of Lotka, other extremal thermodynamic principles have been proposed, including: minimum specific dissipation (Prigogine & Wiame 1946), minimum specific power (Margalef 1963), maximum exergy (Jorgensem & Mejer 1979), maximum entropy production (Ulanowicz & Hannon 1987; Schneider & Kay 1994), and maximum empower (Odum 1988), a variation of Lotka's maximum power principle. All of these thermodynamic optimization hypotheses share a common structure: they all state a given thermodynamic property of actual ecological systems is a maximum or minimum among all other available systems states. Although it has been show that some of the posed thermodynamic principles can be complementary (Fath et al. 2001), others, such as the minimum specific dissipation and maximum entropy principles, cannot be fulfilled at the same time for a given system.

Even though research related to thermodynamic and ecological systems have been an active area over the past 60 years, the ability to test whether ecological systems are indeed subject to thermodynamic optimization have remained a rather elusive subject. The modest amount of empirical evidence is likely to be the main reason behind the limited further development of a thermodynamic ecological theory: the majority of the published works dealing with thermodynamic optimization in ecological systems have been focused on theoretical considerations (e.g. (Bosatta & Göran 2002; Sabater 2006; Dewar 2010). Among papers dealing with experimental data, many have showed agreement between observations and what is expected based on thermodynamic optimization principles (e.g. Jørgensen et al. 2002; Aoki 2006 ; see Jorgensen & Fath 2004 for a review), and only few papers have tested predictions generated from thermodynamic optimization hypotheses (Ludovisi 2004; Cai et al. 2006; Meysman & Bruers 2007; DeLong 2008). This is indeed not surprising: we still lack a general framework for deriving predictions from thermodynamic optimization hypothesis on actual ecological systems.

In this paper we present a novel approach that allows testing particular thermodynamic optimization hypotheses on actual systems. A key point of our work is the consideration of the main role that constraints play in thermodynamic optimization hypotheses: the explicit consideration of general constraints makes it possible to generate predictions in the form of probability distributions. In the first section of the work we state the main bases of the methodology. In the second part of the paper we present the application of the novel method to a general and well known animal population model - the Leslie matrix -, showing how it can be used to derive predictions to test thermodynamic optimization in life history traits. We finish discussing the relevance of the presented results in the context of the current state of ecology, and implications for the further development of a thermodynamic ecological theory.

**The basis of a general methodology to test thermodynamic optimization in ecological systems**

At the core of our approach to test thermodynamic optimization in ecological systems is the notion that optimization phenomena consist not only of an hypothesized extremum principle, but also the constraints that set the frontiers under which the optimization can take place. The optimization of a thermodynamical system requires by definition a number of constraints that set limits to the variable being minimized or maximized. If an ecological system is indeed subject to thermodynamic optimization, the actual system state will be some of the optimum available states given the operating constraints.

Even though constraints are frequently recognized as a central issue regarding thermodynamic and ecology (Kleidon et al. 2010; Volk & Pauluis 2010), they have received scarce attention in research on thermodynamic optimization in ecological systems (although see Dewar et al. 2009). If a thermodynamic property of an ecological system is indeed under optimization, this knowledge would be a necessary but not sufficient condition to predict the actual state of the system: it would also be necessary to know the constraints that determine the available states of the system among which the optimum state was selected. The methodology we propose takes as the hypothesis both the thermodynamic optimization principle and general constraints that limit the optimization process.

The general methodology involves 3 main steps: i) formulation of a mathematical model of the system parametrized with experimental data ; ii) formulation of general constraints that limit the thermodynamic optimization ; and iii) a null hypothesis generation based on both the system model and operating constraints. Each part of the methodological is subsequently briefly described, and a schematic summary is presented in Fig. 1.

*Mathematical model of the system*

The input of the method is a mathematical model of the ecological system under study, which should summarize current biological knowledge about the system by means of both model definition and by parametrization of the model with experimental data. The mathematical model should be of a sufficient degree of detail in order to allow the calculation of thermodynamic properties such as total energy flux, dissipation, and so on, as functions of the parameters of the model. For a model with $k$ parameters, the actual system state will be represented by a point in the $k$-dimensional parameter space, or more precisely a small portion of the space due to variance in parameters.

*Constraints formulation and their relationship with the actual system state*

A set of general constraints expressed as a function of variables or parameters of the model can be posed for the formulated model of the system under study. These constraints are hypothesized to limit the thermodynamic optimization phenomena (we will refer to this at the end of this section). While the actual system state is represented by a point in the parameter space, each constraint will determine a subspace of the parameter space of the model. The parameter subspace determined by a constraint is simply the set of all points in the parameter space which make the system model compatible with the constraints (i.e. the alternative available states). When considering a set of constraints, the associated subspace will simply be the intersection of all subspaces determined by each constraint alone (Fig. 1).

The actual system state will be compatible with any constraints whatsoever that might operate upon it, since the by definition if constraints do really apply the system will obey them. Therefore, the actual system state contains information regarding potential constraint to the thermodynamic optimization. There are two types of constraints that can be in general posed (Biegler 2010). Equality constraints take precise values, and therefore we can estimate the value of any hypothetical equality constraint expressed as a function of the model parameters and variables. Inequality constraints, on the other hand, impose minimum or maximum limits to the system, which can be independently determined by knowledge of the surrounding of the system. However, if no data is available on inequality constraints values, the actual system state still gives us information about these kinds of constraints, since the value of inequality constraints must take at least as extreme values as observed in the actual system. In summary, for any constraints that are hypothesized to act upon the thermodynamic optimization process, knowing the actual system state give us some information on the values of these constraints.

*Deriving predictions*

We propose to estimate the probability density function (PDF) of the thermodynamic property that is hypothesized to be under optimization from the available parameter subspace of the system's model determined by all the constraints. Consider a particular thermodynamic principle and a set of constraints that is (Stauffer 2008). Based on the nature of the posed constraints and the knowledge we have of them, there are two possible ways in which the outlined procedure can be applied. If all constraints are equality ones, or if independent estimates of all inequality constraints are available, the application of the described scheme is straightforward, and allows to test whether there is evidence of thermodynamic optimization under the hypothesized constraints. If, however, independent estimates of inequality constraints are unknown, which is a common situation, a restricted hypothesis testing can still be applied. If no information is available, a portion of the parameter subspace determined by an inequality constraint can be determined based on the actual system state. However, inferences made using this parameter subspace will be biased, since constraints can be correlated to the property under optimization. However, if an inequality constraint is treated as an equality constraint which takes the value corresponding to the actual system state, an unbiased parameter subspace with respect of the constraint can be obtained. The parameter subspace defined in this way will be a portion of the "true" available parameter subspace determined by the inequality constraint. Therefore, if the optimization actually occurs, the actual system state will be an optimum of this restricted subspace, since a global optimum is also always a local optimum. Although an independent estimate of an inequality constraint is obviously better, in the absence of such information the prior procedure allows to perform a test a thermodynamic optimization hypothesis in a restricted way.

*Constraints as necessary accessory hypothesis*

By explicitly formulating constraints we can make predictions to test thermodynamic optimization, but at the same time constraints become in this way part of the hypothesis under testing. In other words, constraints act as accessory hypotheses to the thermodynamic optimization principle. When testing optimization in a given system, if the hypothesis is rejected, this will indicate that either the thermodynamic optimization, the formulated constraints, or both, do not occur. Hence, it would be possible to change the optimization principle or the constraints which is hypothesized to be subject to. Thus, a new challenge appears: which constraints to consider and in what order? We suggest that a minimum number of general constraints should be added at the start of the analysis of a given system. A minimal set of constraints would be those necessary to assure the model behaves with biological realism. If no evidence of optimization is found, further constraints could be added.

**Application to a general animal population model**

We applied the described methodology to a general age-structured population model parametrized using physiological data from the literature to show how it allows to derive predictions to test whether actual age-specific survival and fecundity rates are consistent with either the minimum specific dissipation or the maximum entropy production principle. The issue of optimal life histories has been extensively treated in the ecological literature from various perspectives (e.g. Charnov et al. 2001; Bonsall & Mangel 2004; Brown & Sibly 2006). Optimal life histories have also been studied using concepts and methods related with thermodynamics. For example, Demetrius developed a theory of optimal life histories based on statistical mechanics considerations (Demetrius 1974; Demetrius et al. 2009). We are not aware, however, of any theoretical nor experimental study that attempted to establish whether actual life-history traits could yield optimal thermodynamic properties, such as dissipation rate or total energy flux.

The application of the general methodology to the posed example requires various steps. First, a population model that allows the expression of main energy fluxes of the population as function of parameters of the model is required. Secondly, some general constraints expressed as function of parameters and/or variables of the model are needed, since they will determine alternative available states. Lastly, it is necessary to estimate the PDF of the thermodynamic property that is hypothesized to be optimum by sampling on the available alternative states. All these steps are briefly described below, and finally the generated predictions are illustrated.

*Animal population model*

We used the Leslie matrix population model (LMM), a simple and very well known age-structured model (Leslie 1945; Caswell 2001). The LMM has been extensively used in wild and laboratory population studies (e.g. Groenendael et al. 1988; Heppell et al. 2000; Ezard et al. 2008) and has also been subject to thorough theoretical analyses (Cull & Vogt 1973; Hearon 1976; Conlisk 1988; Gosselin & Lebreton 2009). The LMM has a discrete age-structured, with time as a discrete variable, and has only two type of parameters: survival and fecundity rates. The discrete age-structure of the LMM takes no assumption regarding the relationship of survival rates of different age classes, which is desirable since we did not intended to further constrain the available parameter space with assumptions of specifics functions relating survival and fecundity with age.

Computation of relevant thermodynamic properties of an animal population such as total energy flux or dissipated energy requires a dynamic model, in this case the LMM, since it allows to determine the stable structure of the population from survival and fecundity rates alone (Caswell 2001). However, to fully characterize population energy flux it is also necessary to know how energy flux varies with age. Since body size is the main trait that influence energy flux of individuals (Peters 1983; Gillooly et al. 2001) estimation of whole population energy fluxes is possible if ontogenetic growth as well as parameters related with the energetic cost of reproduction are known.

Data regarding energetic physiology are more abundant for mammals than for any other taxa, and therefore for our example we took general parameters for mammals from published works that broadly correspond to a mammal with a body size of 100 g. A summary of the terms and parameters values used in the animal population model are summarized in Table 1. A general growth model (West et al. 2001; Moses et al. 2008) was used to model variation of body size with age. Parameters of metabolic scaling were taken from Gillooly et al. (2001). Maximum life-span were estimated from general regressions for mammals to be of 1.3 years. We considered a number of 10 different age classes since this is about the commonly used number of age classes used in LMM. Sexual maturity was considered to occur at the age class with 70 % of maximum body size, which according to the growth models occurred at age class 5. A same maximum possible fecundity was considered for all reproducing age classes as this is the most common pattern in species with determinate growth. Population energy flux was modeled taking into account energy outputs of the population, and included heat dissipated due to metabolism, mass lost by mortality, and energetic costs of reproduction. The mathematical expression of energy flux is presented in the next section of the work.

*Formulation of constraints*

We considered two general constraints to illustrate how predictions are derived from both the hypothesized constraints and the mathematical model of the system under study. The first type of constraints we considered was a dynamic one. Imposing a constraint on population dynamics implies the actual state of the system will only be compared with other alternative states with a similar dynamic behavior. Among possible dynamic constraints, the steady state is the most common assumption in ecological models. The dynamic behavior of the LMM depends on survival and fecundity rates alone, and the steady state constrain is mathematically equivalent to a dominant eigenvalue of the projection matrix equal to 1 (Caswell 2001). A steady state constraint selects alternative states in which the combination of survival and fecundity rates produces a non-growing population, and impose to survival and fecundity rates the following mathematical condition (Cull & Vogt 1973):

$$1 = f_1 + \sum_{i=2}^{k} f_i \prod_{j=1}^{i-1} s_j$$

where *k* is the number of age-classes of the LMM.

The second constraint we considered was the total energy flux (TEF). TEF is also a very general constraint, since the rate of energy supply is always limited in any possible situation. As an inequality constraint, the TEF of a system can take up to a certain maximum value depending on the surroundings of the systems. The limit value of TEF cannot thus be estimated by the actual system state itself, although we know that TEF of the actual system will lower than that limit maximum value. We are here interested in generating predictions to test optimization of the dissipation rate. In the absence of an independent estimation of its maximum possible value, which is a common case, TEF could be treated as an equality constraint as previously described in the prior section of the work. Imposing TEF as an equality constraint implies that the actual state of the system will only be compared with alternative states with a similar TEF value. Thus, variation of dissipation rate among alternative states will not depend on a different TEF, but only on the relative amount of the energy flux that is dissipated. In this way, the hypothesis under testing will be regarding the proportion of energy that is dissipated relative to TEF rather than the absolute amount of energy dissipated. This reflects a fundamental fact we have previously pointed out: the hypothesis under testing will be composed by both constraints and the optimization principle. Given the prior existence of the steady state constraint, the mathematical expression of TEF for the LMM of an animal population is:

$$TEF = a\sum_{i=1}^{k} w_i^b n_i^* + c\sum_{i=1}^{k} (1 - s_i) w_i n_i^* + cd\sum_{i=1}^{k} f_i n_i^* + c\sum_{i=1}^{k} w_i n_i^* (1 - 1/\lambda)$$

While the steady state constraint have a precise value by definition, i.e. a dominant eigenvalue equal to 1, the value of the TEF constraint do not have a predefined value. TEF was set to 7.3 $10^{-4}$ W $kg^{-1}$, which is an intermediate value for the animal population model at steady state parametrized as previously described. According to this TEF formulation, the dissipation rate was estimated as:

$$D = a\sum_{i=1}^{k} w_i^b n_i + e\,c\sum_{i=1}^{k} f_i n_i w_1$$

*Generation of predictions*

Along with model definition and constraint formulation, the last step and core of the proposed methodology is to obtain an unbiased sample of the available alternative states determined by the hypothesized constraints. Details of the mathematical procedure followed to obtain an unbiased sample of the available parameter subspace are described in the Appendix. The effect of constraints on the available parameter subspace can be easily observed in Fig. 2, which shows how it is further reduced by each constraint that is added. An unbiased sample of the available parameter subspace determined by all constraints is straightforwardly obtained by simply computing the intersection between available states determined by each constraint.

The probability distribution of the dissipation rate was estimated using the obtained sample of the available parameter subspace (Fig. 3). The probability distribution constitutes a prediction of how dissipation rate would be expected to be if no optimization is took place under the hypothesized constraints, and thus it acts as a null model. The probability distribution we obtained here using general parameters from the literature (Table 1) showed a left-skewed distribution, indicating a rather step limit to low dissipation rate values, whereas a relatively long right tail indicates a more broad limit towards the upper value of dissipation rate.

Alternative available states of the system model are characterized by points belonging to an hypervolume of survival and fecundity rates. We therefore projected available states onto planes formed by survival rates, and coded dissipation rate as gray intensity to allow visualization of the dissipation rate of the generated alternative states. Interestingly, a pattern between dissipation rate and the value of survival rates was revealed (Fig. 4), showing that configurations with similar survival rates have similar dissipation rate.

**Discussion**

Schneider & Kay (1994) stated that a thermodynamically based theory of ecology holds the promise of propelling ecology from a rather descriptive to a predictive science. Although certainly not few ecologists might disagree with such a bold statement, probably all of them would agree that the true value of thermodynamic optimization in ecological systems should be derived by evaluating its explanatory power, i.e. comparing predictions with empiric results. We have presented a novel methodology that constitutes the first general approach to test thermodynamic optimization hypotheses in ecological systems that we are aware of. Almost a century after Lotka (1922a) first proposed the maximum power principle, the issue of whether ecological systems are subject to thermodynamic optimization remains controversial. The development of a thermodynamic view of ecology would most certainly benefit from empiric results supporting or rejecting optimization hypothesis. However, empiric testing of thermodynamic optimization has been challenged by the lack of a general methodology for making predictions for specific ecological systems. DeLong (2008) stressed the importance of generating testable predictions about real biological phenomenon from thermodynamic optimization hypothesis. When competing hypothesis exist, which is the case of thermodynamic optimization in ecology, we might add that it is of fundamental importance to test multiple hypothesis at once to determine which one predicts better empirical results. Here we have shown how a simple yet powerful technique allows deriving specific predictions to test thermodynamic optimization hypothesis in real biological system.

The generality of the methodological approach presented in this work makes possible to apply the developed scheme to ecological systems of different hierarchy and complexity in a similar manner as shown, including communities and ecosystems. Application to systems of different hierarchy would make possible to asses whether thermodynamic optimization occurs at one particular hierarchical level or it is a general property of ecological systems as many authors hypothesize. Although applying the methodology to systems more complex than a single population is a simple and straightforward idea, the numerical methods involved could get computationally intensive as system models get more complex, and the efficiency of numerical methods used to obtain the sample of the available parameter subspace might be critical to keep computing times reasonable. Numerical methods with a general applicability to obtain an unbiased sample of the available parameter subspace in complex models to allow testing thermodynamic optimization irrespectively of mathematical formulation of the model or constraints would certainly be useful.

We were able to derive for the first time a general method to make accurate predictions of thermodynamic optimization in ecological systems by acknowledging the main role that constraints play in any optimization process. Although constraints are a necessary component of any optimization process, little attention have been paid to constraints in research on thermodynamic optimization in ecological systems. An important consequence of the novel methodology we have presented is that it allows to test not only the existence of a thermodynamic optimization, but also to which constraints it is subject to. The inclusion of constraints as accessory hypothesis might be consider by some a further complication of the already considered by most ecologists complex thermodynamic approach. In contrast, the explicit inclusion of the constraints enabled us to formulate predictions that would allow to test thermodynamic optimization hypothesis on specific experimental populations. As early noted by Lotka (1922a), it is not a simple matter to define and formalize the constraints to which systems are subject, and that would restrict the tendency imposed by an optimization principle. We are aware of the challenge of posing constraints with biological meaning and that are not merely mathematical function. However, we believe this could help to clarify the relationship of the thermodynamic optimization approach with ecological theory, bringing more biology to the matter instead of leaving biology aside. In a way, a price is paid by explicitly considering hypothetical constraints, but this enable a predictive power otherwise unattainable, and we thus consider that rewards greatly outwards costs.

Published works dealing with thermodynamic optimization of ecological systems have been focused almost exclusively on the community and ecosystem level. Even though populations are the simplest ecological systems beyond individuals, we are unaware of any work that has tested thermodynamic optimization hypotheses on populations. By mains of the develop methodology we were able to generate predictions to test whether there is evidence that actual survival and fecundity rates of an animal population are not a random set of the possible states defined by the posed constraints, but instead actual rates make the population an optimum state with respect of the thermodynamic property of interest. Experimental data allowing to characterize in detail energy fluxes in populations is by far more accessible than at the community or ecosystem level, and therefore the thermodynamic study of populations could ease the generation of empiric results aimed to bring new insights about thermodynamic optimization in ecology.

**Table 1.** Glossary of terms and values of parameters used in the animal population model

| Term | Description | Value |
|---|---|---|
| $s_i$ | Survival rate of age-class i | Random, subject to constraints |
| $f_i$ | Fecundity rate of age-class i | Random, subject to constraints |
| $n_i$ | Abundance of age-class i of the stable age distribution | Depends on $s_i$ and $f_i$, subject to constraints |
| $w_i$ | Body mass of of age-class i | Calculated from growth model for each age-class |
| a | Normalization constant | $1.07 \times 10^{-6}$ W.kg$^{4/3}$ |
| b | Metabolic scaling exponent | 0,75 |
| c | Energy content of biomass | $7.73 \times 10^{6}$ J kg$^{-1}$ |
| d | Energetic costs of reproduction excluding and relative to energy content of produced offspring | 13.2 |
| e | Energetic costs of reproduction dissipated as heat relative to energy content of produced offspring | 10.4 |
| $\lambda$ | Dominant eigenvalue of the LMM | Constrained, set to 1 |
| TEF | Population total energy flux | Constrained, set to $7.3 \times 10^{-4}$ W kg$^{-1}$ |

**Figures**

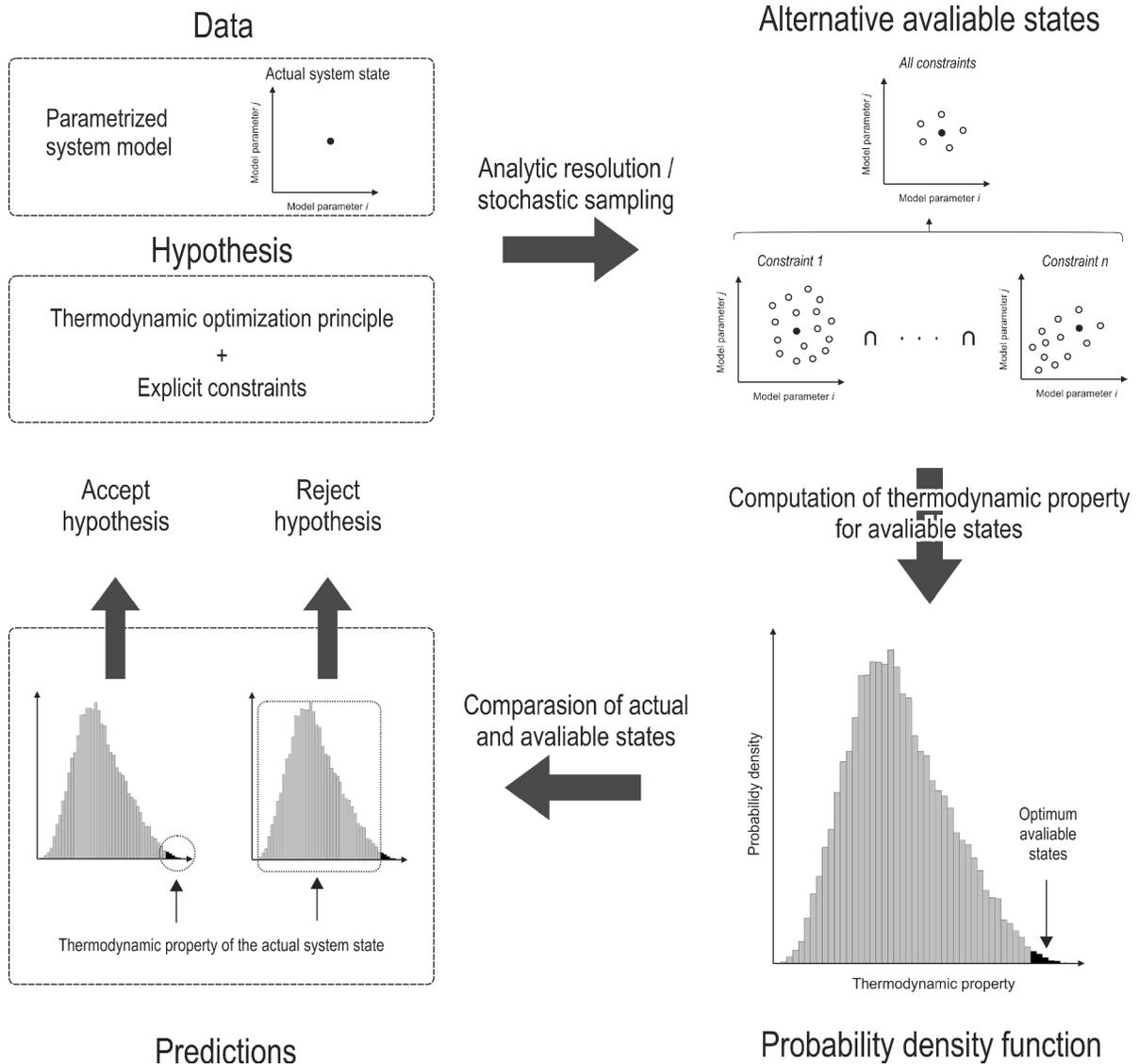

Figure 1. Schematic diagram of the main steps for testing thermodynamic optimization hypotheses of the methodology presented in this work. In the hypothetical example shown, the thermodynamic principle under testing states a thermodynamic property is maximized, thus optimum available are those that present maximum possible values.

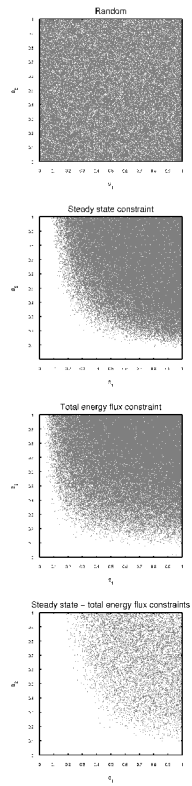

Figure 2. Sequential application of constraints to parameters of the Leslie matrix model of an animal population shown by the projection of alternative available states on the plane formed by the first two survival rates.

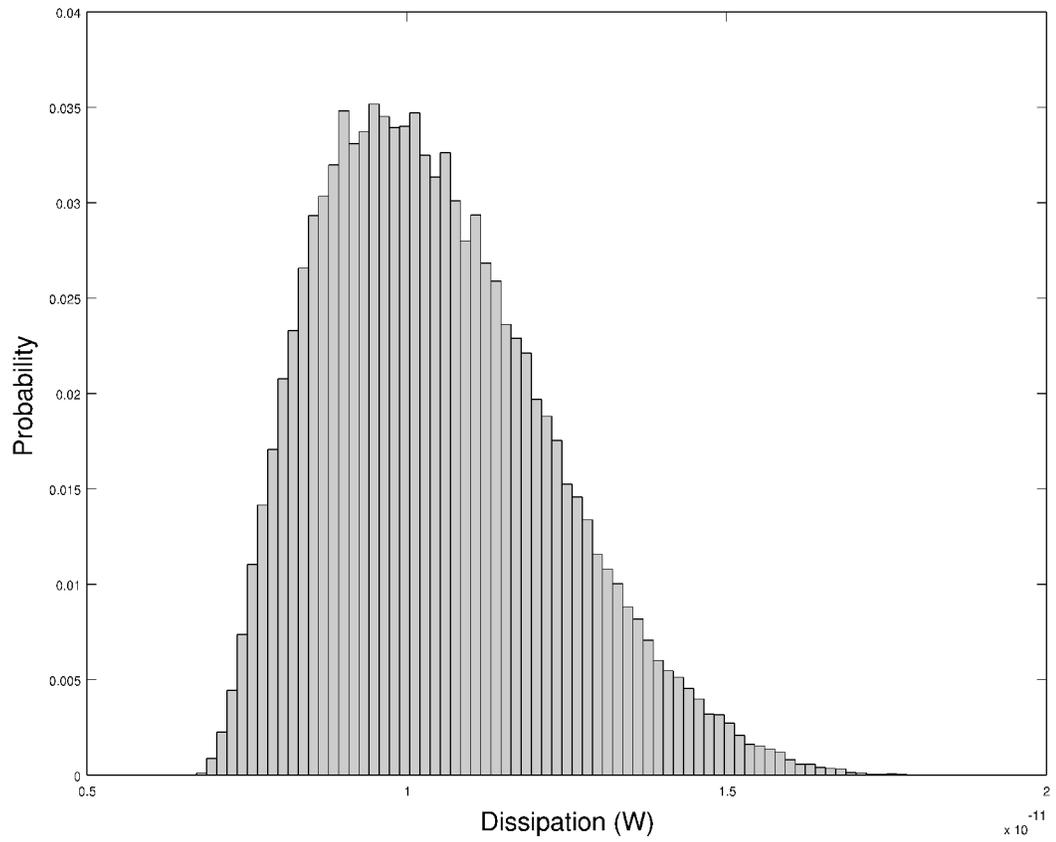

Figure 3. Histogram showing the probability of a dissipation rate range of random alternative states compatible with the steady state and total energy flux constraints.

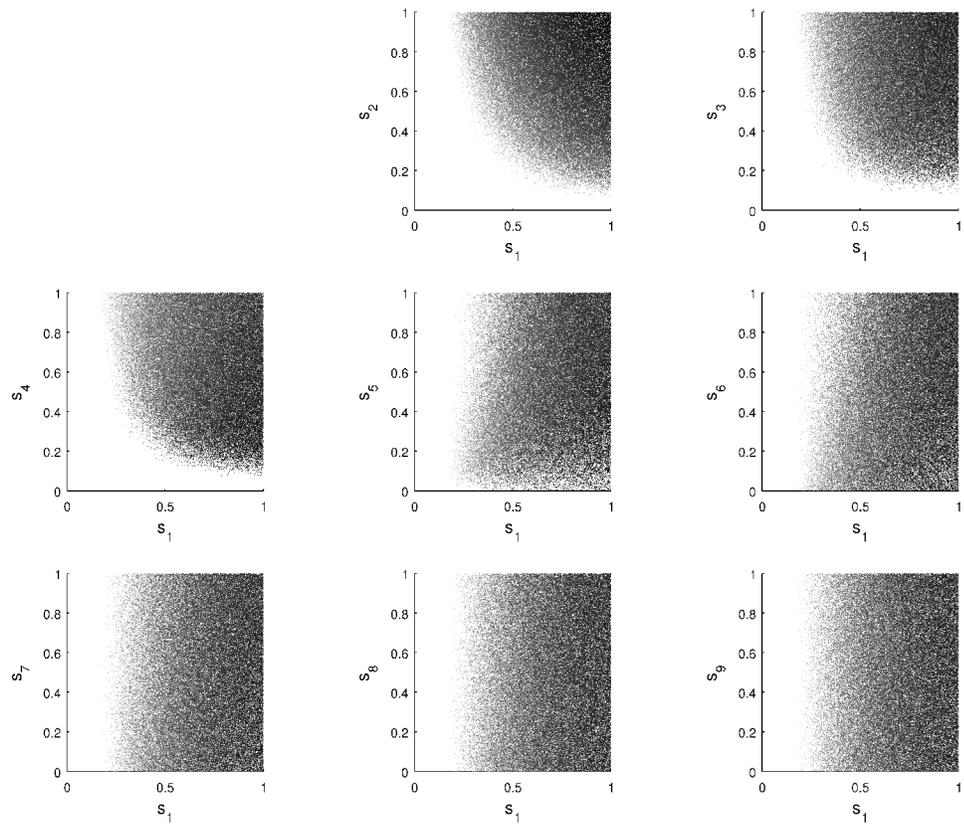

Figure 4. Dissipation rate of the alternative states of the Leslie Matrix population model compatible with constraints, coded as gray values (darker = higher), reveals a pattern between dissipation and survival rates.

**Apendix**

We obtained a sample of the available alternative states as follows. First, we obtained sets of survival and fecundity rates that fulfill the steady state constraint starting with random numbers. We used the equation of the steady state constraint to derive equivalent conditions on each parameter of the Leslie matrix model (LMM) as a function of all other parameters. For a given random set of parameters of the LMM $\{s_1,...,s_k,f_1,...,f_k\}$, one could yield a set that fulfill the steady state constraint by modifying a single survival or fecundity rate. We therefore used the mathematical conditions to evaluate, for each $s_i$ and $f_i$, if a valid value (i.e. $s_i \in (0,1)$ and $f_i \in (0, F_{max})$) existed that would make the random set fulfill the steady state condition:

$$s_i^* = \frac{s_i}{\sum_{j=i+1}^{k} f_j c_j} \left(1 - \sum_{j=1}^{i} f_j c_j\right)$$

$$f_i^* = \frac{1 - \sum_{1}^{k} f_j c_j + f_i c_i}{c_i}$$

where

$$c_1 = 1$$

$$c_i = \prod_{1}^{i-1} s_j, i > 1$$

This probed to be a very efficient way of randomly generating configurations at steady state. Secondly, we used the configurations at steady state to compute their total energy flux, and evaluated whether they fulfill the total energy flux constraint by an error <0.0001%.